\documentstyle[epsfig]{mn2e}

\def\etal{{\rm et all. }}

\def\kpc{{\ h^{-1} \ \rm kpc}}
\def\kms{{\rm km \ s^{-1}}}

\title{Effects of galaxy interactions in different environments}

\author[M. Sol Alonso et al.]{M. Sol Alonso $^{1,2}$,
Diego G. Lambas $^{1,4}$,
Patricia Tissera $^{1,3}$ and
Georgina Coldwell $^{1,4}$\\
$^1$ Consejo Nacional de Investigaciones Cient\'{\i}ficas
y T\'ecnicas.\\
$^{2}$ Complejo Astron\'omico El Leoncito, Argentina \\
$^{3}$ Instituto de Astronom\'{\i}a
y F\'{\i}sica del Espacio, Argentina.\\
$^4$ Observatorio Astron\'omico
de la Universidad Nacional de C\'ordoba,  Argentina.\\
}

\date{\today}

\begin{document}
\pagerange{\pageref{firstpage}--\pageref{lastpage}}

\maketitle

\label{firstpage}

\begin{abstract}

We analyse star formation rates derived from photometric and spectroscopic
data of galaxies in pairs in different environments using the
2dF Galaxy Redshift Survey (2dFGRS) and the  Sloan Digital Sky Survey (SDSS).
The two samples comprise several thousand pairs,
suitable to explore into detail the dependence of star formation activity in pairs on
orbital parameters and global environment. We use the projected galaxy density derived
from the fifth brightest neighbour of each galaxy, with a convenient luminosity threshold
to characterise environment in both surveys in a consistent way.
Star formation activity is derived through the $\eta$ parameter in 2dFGRS and through
the star formation rate normalised to the total mass in stars, $SFR/M^*$, given
by Brinchmann et al. (2004) in the second data release SDSS-DR2.
For both galaxy pair catalogs, the star formation birth rate parameter is a strong 
function of the global environment and orbital parameters. Our analysis on SDSS pairs confirms previous
results found with the  2dFGRS where suitable thresholds for the star formation activity 
induced by interactions are estimated at a projected distance $r_{\rm p} = 100 \kpc$  
and a relative velocity $\Delta V = 350$ km $s^{-1}$.
We observe that galaxy interactions are more effective at triggering important star
formation activity in low and moderate density environments with respect to the control sample of
galaxies without a close companion.
Although close pairs have a larger fraction of actively star-forming galaxies, 
 they also exhibit a greater fraction of red galaxies with respect to
those systems without a close companion, an effect that may indicate that
dust stirred up during encounters could affect colours and, partially, obscure tidally-induced
star formation.

\end{abstract}


\section{Introduction}

The current cosmological paradigm for 
structure formation postulates that galaxies formed by hierarchical aggregation.
In this scenario, galaxy interactions and mergers play a principal role being a possible
efficient mechanism to modify the mass distribution and  trigger star formation activity.
It was Toomre \& Toomre (1972) who first showed that tidal interactions can transform
spiral and irregular galaxies into bulge ellipticals and S0s.
Later on, more sophisticated numerical simulations (e.g. Barnes \& Hernquist 1996; 
Mihos \& Hernquist 1996) showed that the gas component might experience
torques produced by the companion with a subsequent increase of gas density which 
triggers star-bursts during the orbital decay phase of the accreting satellite. 
Tidally induced gas inflows fuel such starbursts.
Cosmological hydrodynamical simulations confirmed these results within a hierarchical
scenario (Tissera 2000; Tissera et al. 2001).

Several observational works showed that  mergers and interactions of galaxies
affect star formation activity in galaxies in the Local Universe (e.g., Larson \& Tinsley 1978;
Donzelli \& Pastoriza 1997; Barton, Geller \& Kenyon 2000; Petrosian 2002) and finding clear signs that
the number of interacting systems increased with redshift 
(see for instance Le Frevre et al. 2000).
However, it is still unknown the impact of mergers and interactions on the life of galaxies and
consequently, the level of agreement with the predictions of hierarchical clusterings.

The study of galaxies in pairs provides useful insights in the nature of 
interactions.
Kennicutt et al. (1987) analysis of close pairs yield a general trend for enhanced star 
formation and nuclear activity although with  a wide dispersion about the mean.
Zepf \& Koo (1989) studied close pairs of faint galaxies, 
separated by less than 4.5 arcsec , finding that although the  colours  of some galaxies
correspond to recent episodes of star formation the overall colour distribution was 
very similar to that of field galaxies. 
Yee \& Ellingson (1995) and Patton et al. (1997) found no significant differences
between the mean properties of isolated galaxies and galaxies in pairs, although those 
which appear to be undergoing interactions or mergers had strong emission lines and 
blue rest-frame colours.

Barton et al. (2000) first 
showed the existence of a clear correlation between the  relative and velocity separations 
and  star formation activity in galaxy pairs in the field. 
Recently,  Lambas et al. (2003, hereafter Paper I) studied  an order of magnitude
larger pair catalog statistically confirming the effects of interactions on
the star formation activity of galaxies,  based on a comparative analysis
of the properties of  isolated galaxies and galaxies in pairs selected from the 2dFGRS
public release data.
More recently, Alonso et al. (2004, hereafter Paper II) extended these investigations to  
high density environments: groups and clusters.  Results from Paper II
 indicated that  pairs in groups were systematically redder and with a lower present-day
star formation activity than other galaxy members, except for 
 galaxy 
pairs with relative separation $r_{\rm p} < 15 \kpc $ which  showed significantly  
higher star formation activity in comparison to galaxy members without a close companion.

Regarding the dependence of star formation activity on environment, 
Mart\'{\i}nez et al. (2002) and Dom\'{\i}nguez et al. (2002) analysed the relative
fractions of passively star-forming galaxies in high density regions corresponding
to groups of galaxies extracted from the 2dFGRS. 
A similar analysis carried out by Gomes et al. (2003) and Balogh et al. (2004)  in
the SDSS and 2dFGRS  surveys, also gave a clear indication of a strong dependence of star 
formation on environment which tend to decrease with increasing density. 
Balogh et al. (2004) found that, at fixed galaxy luminosity, the fraction of the red galaxies 
was a strong function of local density, increasing up to $\sim$ 70 per cent of the 
population in the highest density environments. 
Similar trends were obtained by Hogg et al. (2003), who showed that the red galaxies, 
regardless of luminosity, are found in overdense regions.  
Kauffmann et al. (2004) used a complete sample of galaxies from SDSS, 
to study the structure and star formation activity as a function of  local density and  stellar mass,
finding that the star formation activity was the galaxy property most sensitive to environment with
an strongest dependence for smallest stellar mass systems. These authors also claimed
that mergers could lead to this dependence, although  environment driven processes such
as tidal stripping,  which could remove gas from galaxies quenching their
star formation activity might be important in high density regions
principally for low stellar mass systems.

These works extended the pioneer studies by
Dressler (1980) who analysed the morphology-density relation showing that the star 
formation activity resided preferentially in disc galaxies in lower density regions.
Different physical processes may play a role in driving the morphology-density relation among which
mergers and interactions stands out in a  hierarchical universe.
Even for systems in the field or in groups,  the 
cumulative effects of many weaker encounters (Richstone 1976; Moore et al. 1996) or few merger
events could have imprinted important features in their astrophysical properties.

In this paper we focus on the statistical study of the effects of 
 the presence of a close companion on colours and
star formation activity of galaxies in different environments. For this aim, we use the sample of 
pairs obtained from the 2dFGRS in Paper II  and   construct  a pair catalog from the SDSS-DR2
 which we have added star formation rates and stellar masses  estimated by Brinchmann et al. (2004). 
This paper is structured as follows: Section 2 describes how catalogs are built up and possible
incompleteness and selection effects. Section 3 discusses the dependence of star formation on
orbital parameters and environments. Section 4 deals with colour distributions as a function
of environment. And in Section 5 we summarize the main findings.

\section{Observational Data and Galaxy Pair catalogs}

In this Section we describe the selection procedure of galaxy pairs from both 
SDSS and 2dFGRS surveys. 
We also discuss possible aperture effects in both pair catalogs and incompleteness 
problems of SDSS pairs. 
Regarding the star formation activity, colours and local density, 
we include a description of these parameters.

\subsection{Pairs from the SDSS survey} 

The SDSS (Abazajian et al. 2004) is a photometric and 
spectroscopy survey that will
cover approximately one-quarter of the celestial sphere and collect spectra of more
than one million objects.
The imaging portion of the second release SDSS comprises 3324 
square degrees of sky imaged in five wave-bands
($u$, $g$, $r$, $i$ and $z$) containing
photometric parameters of 53 million objects. 
The main galaxy sample is essentially a magnitude limited spectroscopic sample 
(Petrosian magnitude) \textit{$r_{lim}$}$ < 17.77$, most of galaxies span a
redshift range $0 < z < 0.25$ with a median redshift of 0.1 (Strauss \etal 2002).
Within the survey area, DR2 includes 
spectroscopic data which cover 2627 square degrees with 186240 spectra of galaxies, 
quasars, stars and calibrating blank sky patched.
The spectra are obtained from 3 diameter fibers (7 square degree area) 
projected on the sky and the spectrographs produce data covering 3800-9200 ${\AA}$.
The spectral resolution at $\lambda$ $\sim$ 5000 ${\AA}$ is
$\sim$ 2.5 ${\AA}$ and redshift uncertainties result approximately in 30 km $s^{-1}$.   

We have estimated the 
stellar birth rate parameter, $b =(1-R) t_H (SFR/M*)$, where
$t_H$ is the Hubble time, $R$ is the fraction of the total stellar 
mass initially formed that is return to the interstellar medium over
the lifetime of the galaxy, and $SFR/M*$ is the present star formation 
rate normalised to the total mass in stars given by Brinchmann et al. (2004). 
We use the mean value $R=0.5$ estimated for galaxies 
by Brinchmann et al. (2004).

\begin{figure}
\centerline{\psfig{file=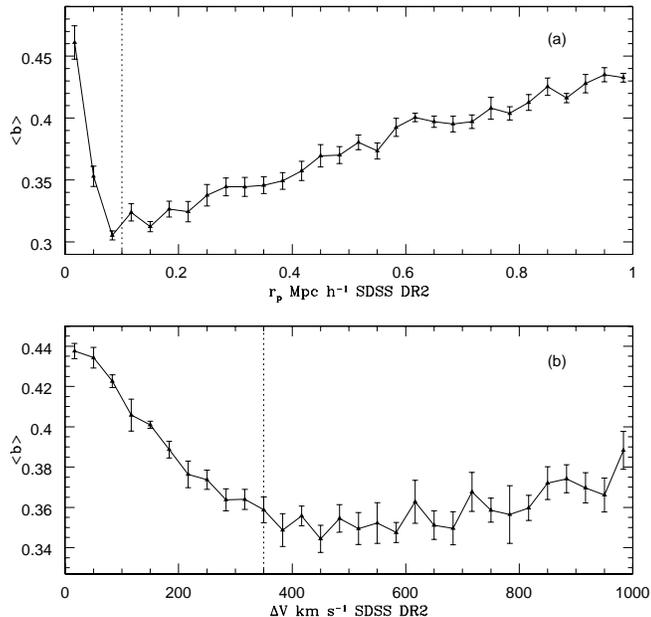,width=9cm,height=9cm}}
\caption{Mean birthrate parameter $b$  as a function of projected separation $r_{\rm p}$ (a)
and relative velocity $\Delta V$ (b) for SDSS-DR2 neighbours.
The dotted vertical lines depict the projected separation and relative velocity thresholds
used for the identification of  galaxy pairs.
Error bars here and in all figures of the paper correspond to uncertainties derived from 
the Bootstrap resampling technique.
}
\label{SFRvec}
\end{figure}

We have considered a redshift range $0.01<z<0.1$ 
in order to avoid strong incompleteness at larger distances as well as
 significant contributions from peculiar velocities at low redshifts.
In order to find suitable limits in projected distance $r_{\rm p}$ and
relative velocity $\Delta V$ to identify galaxy pairs with star formation
enhancement in SDSS we followed the procedure described in Paper I and Paper II.
We firstly analyse neighbours in concentric spheres within $r_{\rm p} < 1$ 
$h^{-1}$ Mpc and $\Delta V < 1000$ km $s^{-1}$, centered at a given galaxy.
For these neighbours, we show in Fig.\ref{SFRvec} the mean birthrate parameter
$<b>$ as a function of projected separation, $r_{\rm p}$, 
and relative velocity, $\Delta V$.
As it can be appreciated from this figure, there is a clear trend for 
the closest neighbours to have higher  star formation activity.
We find a significant enhancement of $<b>$ in galaxies with $r_{\rm p} < 100 \kpc $
and $\Delta V < 350$ km $s^{-1}$ (vertical lines) with respect to the mean
stellar birth rate parameter value of total SDSS-DR2 survey, $<b>= 0.35$.
By imposing these thresholds,
a SDSS-DR2 galaxy pair catalog of 7674 galaxy pairs was built up
which include information on the star formation activity of all its members.

These thresholds  are similar to those inferred from the 2dFGRS 
catalog  confirming that these limits are reliable to select pair galaxies 
with statistically enhanced star formation activity. 
Even more, this good agreement between 2dFGRS and SDSS provides confidence 
on previous results and stimulates new analysis on the nature of star 
formation activity triggered by interactions.

Contamination by active galactic nuclei (AGN) could contribute
to the emission spectral features affecting our interpretation
of star formation activity. In order to analyse this
effect, we cross-correlated the AGNs catalog constructed by Brinchmann et
al. (2004) with our SDSS-DR2 pair catalog. 
We found that  $\approx$ 17$\%$ of galaxies in our total pair sample 
were classified as AGNs. Extracting these AGNs sources, the final SDSS-DR2 galaxy pair catalog 
comprises 6405 pairs.

\subsection{Pairs from the 2dFGRS}

The 2dFGRS is one of the largest present-day spectroscopic survey. 
It includes spectra for 245591 objects with determined redshifts for 221414 
galaxies brighter than a limit magnitude in $b_j$ = 19.45. 
The survey covers an area of approximately 1500 square degrees in three regions:
NGP strip, SGP strip and 100 random fields.  
The strip in the North Galactic Hemisphere has 70000 galaxies and cover $75^0$ x $7.5^0$.
There are approximately 140000 galaxies in the $75^0$ x $15^0$ South Galactic 
Hemisphere strip centered on the South Galactic Pole.  
The 100 random fields scattered over the entire southern region of the APM galaxy survey
outside of the main survey strip comprise 40000 galaxies.

In this paper, we selected pair galaxies following the lines discussed in  Paper I and Paper II.
The pair selection was re-done so that pairs could be identified in all environments
from voids to rich clusters.
We adopted the thresholds $r_{\rm p} = 100 \kpc $ and $\Delta V = 350 $ km $s^{-1}$,
according to our previous works.
The final pair catalog in the 2dFGRS comprises 6067 galaxy pairs. 
We have considered a redshift range $0.01<z<0.10$, but note, that in this case,
 AGNs have not been extracted.

The 2dFGRS provides a spectroscopic classification of the star formation activity 
of galaxies by the $\eta$ parameter obtained from a principal component analysis. 
In Paper I we deduced a linear correlation between $b$ and spectral-type 
index $\eta$ which allowed the estimation of the $b$ parameters for these galaxies.
However, in order to estimate a birth rate parameter for galaxies in the 2dFGRS which 
could be compared to that obtained for SDSS-DR2, we use pair galaxies in common in 
both catalogs to correlate the $b$ parameter derived by the $SFR/M^*$ in SDSS with the $\eta$
index in 2dFGRS. We find that $b(SFR/M^*)$ and $\eta$ define a  linear correlation
of the form $b(\eta)=((0.25 \times \eta)+1.06)-0.35$ (see Fig.\ref{histz}a), except for the lowest
$\eta$  values. However, for the analysis we carry out in this paper, this  underestimation of
$b$ for galaxies with very extreme $\eta$ values does not affect our results.
In all our analysis, we have therefore adopted this $b(\eta)$ relation for 2dFGRS galaxies.

\subsection{Aperture and Incompleteness Effects}

We discuss the possible presence of  
systematics that could bias our star formation rates and pair definition, namely
$aperture$ and $incompleteness$. Fiber angular sizes are 2" and 3" for 2dFGRS and SDSS 
while incompleteness is expected to affect more strongly the SDSS. 
Here we evaluate incompleteness effects for the SDSS by combining the photometric
and spectroscopic catalogs. 

\subsubsection{ Possible Aperture Effects:}
Aperture effects is an important concern because these surveys have observed all
galaxies through a fixed 2" (2dFGRS) or 3" (SDSS) size fibers, smaller
than the average angular size of galaxies.
This can lead to minimize the effects of star formation in large
disc galaxies. Possible effects of aperture bias have been discussed in
some detail by several authors (Baldry et al. 2002; Gomez, et al. 2003;
Balogh et al. 2004; Kauffmann et al. 2004; Brinchmann et al. 2004).

Balogh et al. (2004) analysed the aperture effects in different
environments finding that there was no significant trend of galaxy
size with local density. 
Therefore, the effect of aperture bias depends mainly on the spatial
distribution of star formation across the galaxy with a lack of strong
environment biases.
From galaxies in SDSS, Brinchmann et al. (2004) found that there are a still strong
aperture effects in $SFR/M^*$ for galaxies with log $M^* > $ 10.5. 
This is expected since these galaxies
often have prominent bulges, in which the specific SFR is expected to
be low. This would affect the star formation rate of the most massive
galaxies in our sample, but not the intermediate and low mass systems,
which dominate our statistics of interacting pairs ($\sim$ 70 $\%$).
 
From these analysis we conclude that this potential bias is not likely to have a
large effect on our analysis of star formation in pair galaxies in
different environments.

\subsubsection{Incompleteness Effects}

In order to assess the effects of incompleteness, we built up 
 a subsample of pairs free from incompleteness effects by cross-correlating the 
spectroscopic and the photometric surveys. We explored the fields in the photometric SDSS survey
 around each spectroscopic pair restricted to $m_r=17.5$\footnote{ With this magnitude
restriction we are considering $\approx $ 70 per cent of the total sample}, searching for 
 those galaxy pairs without any extra galaxy companion in the photometric survey
 within a projected distance of $100 \kpc$ at the same limiting magnitude.
By doing so, we obtain a
subsample of pairs that is free from spectroscopic incompleteness bias
and is, therefore, appropriate to test the results for the samples analysed
in the paper against incompleteness effects of the spectroscopic survey.
By comparison between both pair samples (clean and contaminated), we estimated that the spectroscopic 
catalog has an incompleteness of $\approx$ 9.5$\%$. Although this is not a large
fraction, we have examined the possible effects in our analysis in Section 3 
where we conclude that there is not a serious bias in our results.

\subsection{Control Samples} 

In order to unveil the effects of interactions in different environments
we constructed control samples for the  2dFGRS and SDSS pair catalogs
defined by galaxies without a close companion within the adopted separation 
and velocity thresholds.  
By using a Monte Carlo algorithm, for each galaxy pair, we selected 
two other galaxies without a  spectroscopic  companion within $r_{\rm p} < 100 \kpc $ and 
$\Delta V < 350 \ \kms$. 
Moreover, these galaxies were also required to match the observed 
redshift and luminosity distributions of the corresponding pair sample.
In this way they will also share incompleteness effects and any other selection bias
which may depend on redshift and luminosity.
We did not impose other restriction since the purpose of the 
control sample is to take into account any possible  
redshift and luminosity  bias while allowing a confrontation of other 
physical parameters such as colours and star formation activity as a 
function of redshift.

\begin{figure}
\centerline{\psfig{file=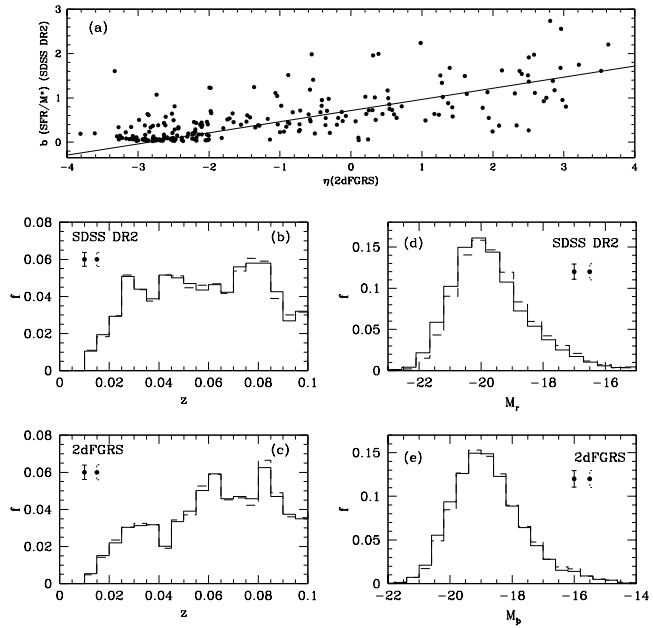,width=9cm,height=9cm}}
\caption{Panel a: $b(SFR/M^*)$ versus $\eta$ for pair galaxies
in common from both catalogs. The solid line represent the linear correlation 
of the form $b(\eta)=((0.25*\eta)+1.06)-0.35$. 
Panels (b) and (c) show the redshift distributions and 
panels (d) and (e) the $M_r$ and $M_{b_j}$ distributions in SDSS and 2dFGRS catalogs,
respectively.
The solid lines correspond to galaxies in  pairs and dashed lines correspond to
galaxies in the control samples. Error bars denote the mean Poisson errors.
}  
\label{histz}
\end{figure}

In Fig.\ref{histz} (panels $b$ and $c$) we show the redshift distributions
for SDSS and 2dFGRS pair samples (solid lines) and their corresponding 
control catalogs (dashed lines).   
Similarly, panels $d$ and $e$ show the $M_r$ and $M_{b_j}$ distributions 
from SDSS and 2dFGRS. It can be appreciated that luminosities and redshift
for both pairs and control samples are similarly behaved in SDSS and 2dFGRS
surveys with very small Poisson errors.

\subsection {Characterizing environment in 2dFGRS and SDSS surveys}
 
\begin{figure}
\centerline{\psfig{file=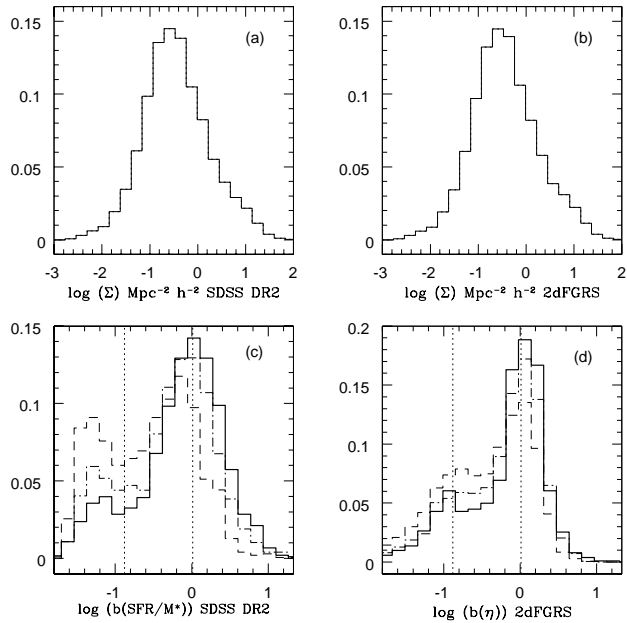,width=9cm,height=9cm}}
\caption{Distribution of $log$ ($\Sigma$) for galaxy pairs in the SDSS (a) and  
2dFGRS (b) surveys.
Distribution of $log (b (SFR/M*))$  and 
$log (b (\eta))$
for pair galaxies in SDSS (a)  and 2dFGRS (b) catalogs in different environments: 
$log$ $\Sigma >$ 0.05 (dashed lines), -0.57 $<$ $log$ $\Sigma <$ 0.05 (dot-dashed lines) 
and $log$ $\Sigma <$ -0.57 (thick solid lines). 
}
\label{histSig}
\end{figure}

The characterization of  the local environment of galaxies 
 is attained by defining a projected local density 
parameter, $\Sigma$.
This parameter  is calculated through the projected distance $d$ 
to the $5^{th}$ nearest neighbour, $\Sigma = 5/(\pi d^2)$.
Neighbours have been chosen to have luminosities above a certain  threshold
and with a radial velocity difference lesser than 1000 km $s^{-1}$.
In a similar way as Balogh et al. (2004), we imposed the condition 
$M_r < -20.5$ to select neighbours in SDSS.
For  the 2dFGRS catalog we estimated a corresponding magnitude limit of  $M_b = -19.3$ by
requiring that galaxy pairs in the common region with the SDSS had a similar
density parameter $\Sigma$ in both catalogs.  
This behaviour is illustrated in Fig.\ref{histSig}(a,b), where we show 
the distribution of the derived values of $\Sigma$ for both pair catalogs.

We  explored if the results obtained in theirs paper could depend on our 
particular choice of local density estimator. 
For that purpose, we also used  the local densities 
 given by Kauffmann et al. (2004). These authors 
estimated the local density  by counting galaxies within cylinders of 2 
Mpc in projected radius and $\pm$ 
500 km $s^{-1}$ in depth, in a complete sample of galaxies from SDSS.
We found a good correlation signal between both density estimators 
indicating  that either of these two definitions 
are adequate to characterize the environment of galaxies.  Also, this comparison has proven that
 incompleteness effects in our sample  are not 
significantly affecting  our  $\Sigma $ estimations. 
The  results discussed in the following sections  have been checked  to be robust against
the particular definition of local environment. Hence 
we claim, that it is environment, 
and not a particular way of characterizing it, that determines the observed trends found in our work.     
In order to analyse in to more detail the dependence of star formation on environment,
in Fig.\ref{histSig}(c,d) we show  the histograms corresponding to the 
birth rate parameter, $log (b)$, obtained from $SFR/M^*$ for 
pairs in the SDSS and from $\eta$ for pairs the 2dFGRS surveys, as a function of
local density.  
We also define three environment classes: low, medium and high, by selecting three suitable 
ranges of $\Sigma$ values in order to have equal number of galaxies in each class. 
These  density thresholds are shown in Table 1.
In order to assess the significance of $\Sigma$ thresholds, we have computed this parameter
for galaxies in 2dF galaxy groups of Merch\'an \& Zandivarez (2002), finding 
$log$ $\Sigma$ values  in the range $0.04$ to $2.02$ in consistency with our present
definition of high density environment ($log$ $\Sigma > 0.05$) .

In Fig.\ref{histSig} (c,d)  we plot the distribution of  birth rate parameter for both 
galaxy pair catalogs segregated in the three defined environment classes.
As it can be clearly appreciated, strong star-forming galaxies tend to avoid 
high density regions. From this figure, the reader can judge how similarly the 
star formation activity as a function of the  local density behaves for the both pair catalogs.
 We claim  that this agreement is good enough to provide a suitable joint analysis of 
the two data sets.

\begin{table}
\center
\caption{Local Density Ranges}
\begin{tabular}{|c c c| }
\hline
Environment & $\Sigma$ Ranges ($Mpc^{-2}$ $h^{-2}$) &  $d$  (Mpc $h^{-1}$)  \\
\hline
\hline
Low          &  $log(\Sigma) < -0.57$         &    2.4         \\
Medium       &  $-0.57 < log(\Sigma) < 0.05 $ &  2.4 $< d <$  1.2  \\
High         &  $log(\Sigma) > 0.05$          &    1.2         \\
\hline
\end{tabular}

{\small  $d$: averaged distance to the $5^{th}$ nearest neighbour 
(see Section 3 for details).}
\end{table}

\section {Dependence of Star Formation on Orbital Parameters and Environment}

In Paper I and Paper II we analysed the star formation activity in pairs in 
two well-segregated environments: field and groups, finding that galaxies in 
close pairs always exhibited enhanced star formation with respect to galaxies 
without a close companion. It is interesting to further investigate the 
effects of environment with a larger sample where 
environment now can be characterized as a more continuous variable from very 
low density regions to very high ones.
Moreover, the possibility of working with two different catalogs will allow us to 
challenge the robustness of the results.

\begin{figure}
\centerline{\psfig{file=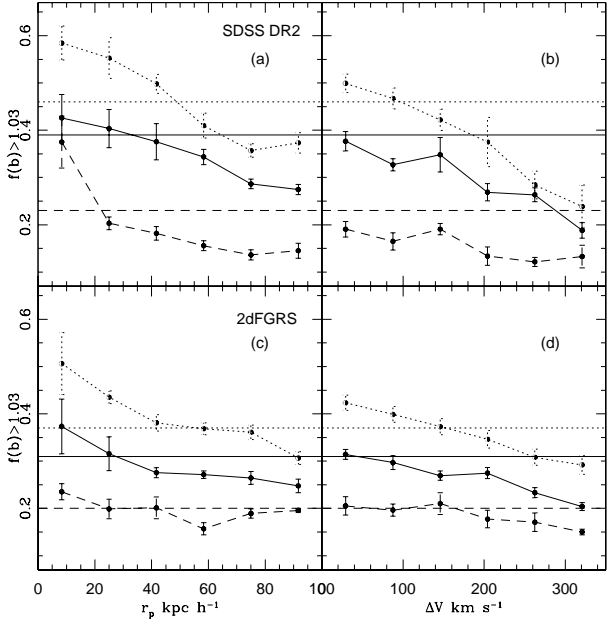,width=9cm,height=9cm}}
\caption{Fraction of strong star-forming galaxies (with $b>1.03$) in the 
SDSS (upper panel) and 2dFGRS (lower panel) pair catalogs 
as a function of the projected distance $r_{\rm p}$ (a, c) and 
relative velocity $\Delta V$ (b, d) for the three different environmental classes:
$log$ $\Sigma >$ 0.05 (dashed lines),      
-0.57 $<$ $log$ $\Sigma <$ 0.05 (solid lines) and $log$ $\Sigma <$ -0.57 (dotted lines). 
The horizontal lines show the corresponding  fractions 
of the control samples.
}
\label{SFRrp}
\end{figure}

We first explored the dependence of the relation between the star formation activity and the 
orbital parameters on environment found in Paper I and II. For this purpose we 
calculated the fraction of strong star-forming galaxies as a 
function of the orbital parameters in pairs and in the control 
samples in both surveys. We adopted the value $b>1.03$ for both 
SDSS and 2dFGRS pair samples since this value define the high star formation 
peak in Fig.\ref{histSig} for the three defined environmental classes.
The results are displayed in Fig.\ref{SFRrp}
from where we  notice a clear increase of the star formation activity for 
smaller $r_{\rm p}$ values, a tendency that is more significant in low 
density environments.
A similar behaviour is observed for the dependence on relative velocity,
$\Delta V$, where again star formation is enhanced with respect to the corresponding fraction 
of the control sample for smaller relative velocities and it
is clearly over the  value of the control samples only in low 
density environments.
Our results are consistent with those derived in close galaxy pairs
by Nikolic et al. (2004), who find the specific star-formation rate to 
decrease at large relative velocity although  
this effect is small compare to that observed with relative separation, $r_p$.   

Therefore, from the analysis of Fig.\ref{SFRrp},
we deduce that galaxy pairs with $r_{\rm p} \approx 50 \kpc $ 
and $\Delta V \approx 200$  km $s^{-1}$ have a statistically 
significant enhancement of star formation activity, regardless of environment.
Hence, we define a close pair subsample formed by  those systems within such limits.

\begin{figure}
\centerline{\psfig{file=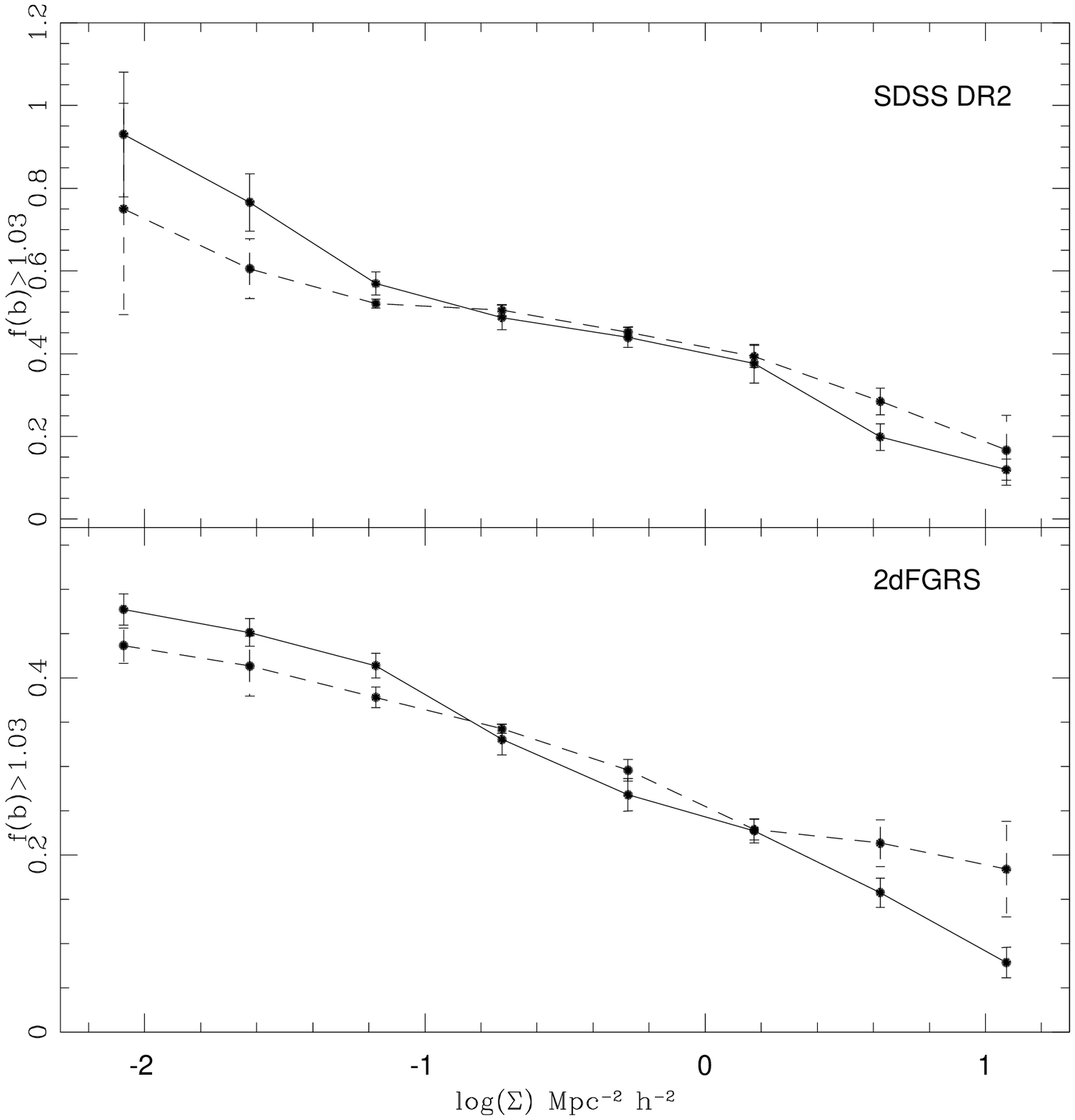,width=9cm,height=9cm}}
\caption{Fraction of strong star-forming galaxies  (with $b>1.03$)
  as a function of $log \Sigma$ in close pairs: 
$r_{\rm p} < 50 \kpc $
$\Delta V < 200$ km $s^{-1}$ (solid lines) and in the control samples (dashed lines)
  in SDSS (upper panel) and 2dFGRS  (lower panel)  surveys. 
}
\label{sfrbsig}
\end{figure}

In order to study into more detail how the fraction of actively star-forming galaxy 
in close pairs varies with environment, we have computed the fraction of strong 
star-forming galaxies as a function of projected local density parameter, $\Sigma$. 
The correlation with $\Sigma$ is illustrated in Fig.\ref{sfrbsig}.
from   where we can see  a strong dependence of the star formation 
activity on local density for the close 
pair samples and their corresponding control ones.
It is clear that close pairs show enhanced star formation activity compared to galaxies in the control 
sample for  densities lower than  $log$ $ \Sigma \simeq -1.0 $. 
For $log$ $\Sigma > 0 $, pairs show a lower star formation activity than 
galaxies without a close companion, a fact already discussed in Paper II.
The transition area is within $log$ $\Sigma \simeq -1$ and $log$ $\Sigma \simeq 0$, where
the fraction of strong star-forming systems seems to be independent of the presence of
a close neighbour. This density range can be associated with that of loose group environments.

As discussed in Paper II  (see also P\'erez et al. 2005 for
a comprehensive analysis of interlopers), interlopers can affect more strongly higher 
density regions
but, the probability to have interlopers is smaller as we take closer pairs.
Moreover, contamination by loose galaxies in the field would tend to increase the mean 
star formation toward that of the control sample. 
Hence, it is unlikely that the behaviour found in Fig.\ref{sfrbsig} 
can be caused by interlopers.

\begin{figure}
\centerline{\psfig{file=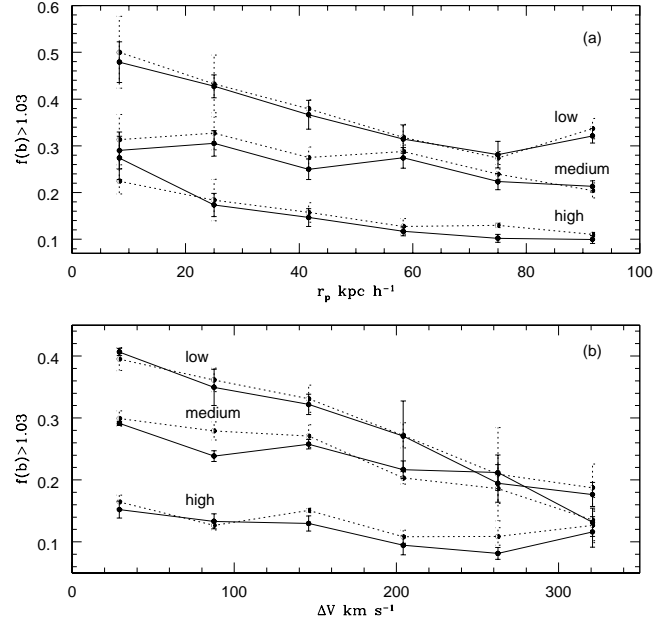,width=9cm,height=9cm}}
\caption{Fraction of galaxies in pairs restricted to $m_r \le 17.5$
with $b>1.03$ for the total  
(solid lines), and clean (dotted lines) samples in SDSS,    
as a function of $r_{\rm p}$ (a), and 
$\Delta V$ (b), for 
high ($log$ $\Sigma >$ 0.05), medium ( -0.57 $<$ $log$ $\Sigma <$ 0.05 )
and low ($log$ $\Sigma <$ -0.57) density environments.  
}
\label{SFRrVcomp}
\end{figure}

As mentioned in subsection 2.3.2, in the case of the SDSS survey, incompleteness in the spectroscopic 
sample could introduce possible systematic bias in our results. In spite of the
low fraction of contamination ($<$ 10 \%) we have studied the star
formation activity for the clean and contaminated subsamples 
($m_r \le 17.5$) from SDSS catalog defined previously.
We calculated the fraction of strong star-forming galaxies ($b >1.03$) as a 
function of projected distance and relative velocity, 
in the three different density environments (Table 1). The results are shown in
Fig.\ref{SFRrVcomp} where it can be seen that the trends are very similar in both samples
within the quoted uncertainties. This test indicates that incompleteness 
in the spectroscopic SDSS survey is not a serious concern for our analysis.

\begin{figure}
\centerline{\psfig{file=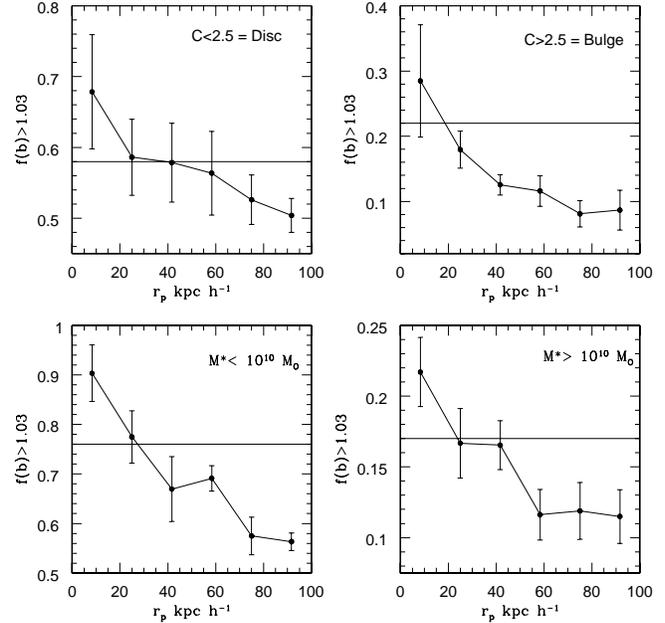,width=9cm,height=9cm}}
\caption{Upper panels: Fraction of galaxies with $b>1.03$ in the SDSS  
pair catalog as a function of $r_{\rm p}$  for large  
$C > 2.5$  or low $C< 2.5$ concentration index values corresponding to bulge or disc 
dominated morphologies, respectively.
Lower panels: Same as in upper panels for subsamples 
with low and high mass in stars ($M^*< 10^{10} M\sun$ and $M^* > 10^{10} M\sun$).
The horizontal lines in the four panels correspond to the  fractions 
of the control samples.
}
\label{bCM}
\end{figure}

Given the strong dependence of star formation activity on morphological
type, we have tested in SDSS if the effects of interactions obtained, depend on
the concentration parameter $C$ (ie. is the ratio of Petrosian 90 \%- 50\% 
r-band light radii) and on stellar mass $M^*$ given by Kauffmann et al. (2004).
Galaxies with a bulge (disc) dominated morphology have $C >  2.5  (C <  2.5)$ values, so we have
analysed the fraction of star-forming galaxies in the pair sample as a
function of projected separation for these two concentration-type galaxies 
separately. The results are shown in Fig.\ref{bCM} where it can be appreciated the increase
of star formation activity at small separations 
for both disc and bulge dominated galaxies. Of course,
disc-dominated galaxies have in general a much larger fraction of star-forming 
galaxies than bulge-dominated systems. But, regarding the
comparison of pair to control samples, the two types of objects behave
in a similar fashion under the presence of a close companion. 
Similarly, galaxies with small mass in stars are more likely to be strong star-forming
 systems than galaxies with large stellar masses. However, the
trend of increasing star formation activity for close relative
separation is observed in a similar fashion for both large and low stellar mass systems.

Hence the fact that we detect lower star formation activity in pairs in the denser regions with 
respect to galaxies without a close companion in the same kind of environment 
suggests that galaxy pairs as bound systems could be more evolved than their 
surrounding loose galaxies.  In fact when the local density increases, 
 the fraction of strong
star-forming galaxies decreases gradually supporting the fact that, as the probability of 
mergers and interactions increased with density, so does the level of evolution 
of those systems that have remained as pairs in these environments.

A comprehensive analysis of these environmental  effects 
requires also a study of extremely low star-forming systems. 
We calculate the fraction of galaxies with low star formation activity in pairs
and in the control samples as those with $b<0.13$ in both SDSS 
and 2dFGRS surveys as a function of projected galaxy 
separation between pair members, $r_{\rm p}$, 
and relative velocity, $\Delta V$, for the same three different environments 
used in Fig.\ref{SFRrp}. This $b= 0.13$ value corresponds to approximately 
the low SFR peak  for the SDSS and the 2dFGRS distributions as 
it can be seen in Fig.\ref{histSig}. 

\begin{figure}
\centerline{\psfig{file=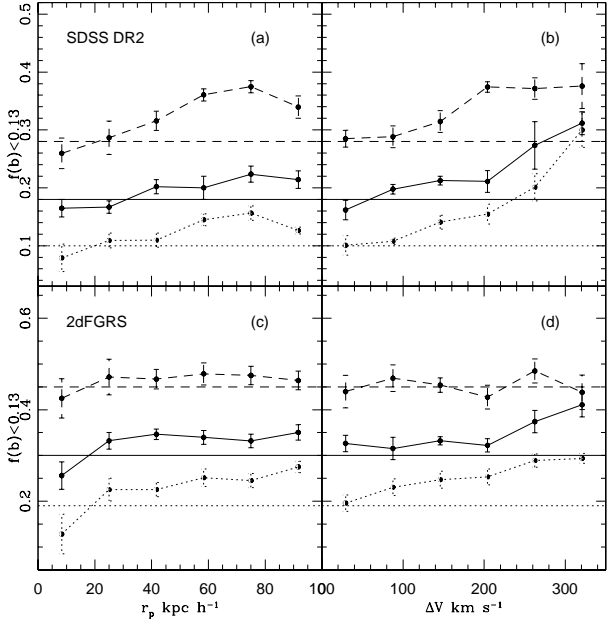,width=9cm,height=9cm}}
\caption{Fraction of low star-forming galaxies (with $b<0.13$) 
as a function of projected distance, $r_{\rm p}$ (a, c), and 
relative velocity, $\Delta V$ (b, d), for three different environmental classes:
$log$ $\Sigma >$ 0.05 (dashed lines),      
-0.57 $<$ $log$ $\Sigma <$ 0.05 (solid lines) and $log$ $\Sigma <$-0.57 
(dotted lines), in the SDSS (upper panel) and 2dFGRS (lower panel) surveys. 
The horizontal lines show the fractions of the corresponding control samples. 
}
\label{SFRrpnonfor}
\end{figure}

From  Fig.\ref{SFRrpnonfor},
we can notice a clear increase of the low star-forming galaxy fraction in  pair  
with larger $r_{\rm p}$ and $\Delta V$ values, a trend that is more significant in 
high density environments. As we mentioned before, the effects of spurious pairs
should be higher for larger relative projected separations and in higher density regions.
Hence, part of the trend could, in principle,  reflect the effects of interlopers.
However, on one hand, for close pairs these effects are not that significant as 
mentioned before and, on the other,  the incorporation of 
interlopers should push the average of pairs toward that of the control
sample while we are actually getting a lower value for close neighbours.

\begin{figure}
\centerline{\psfig{file=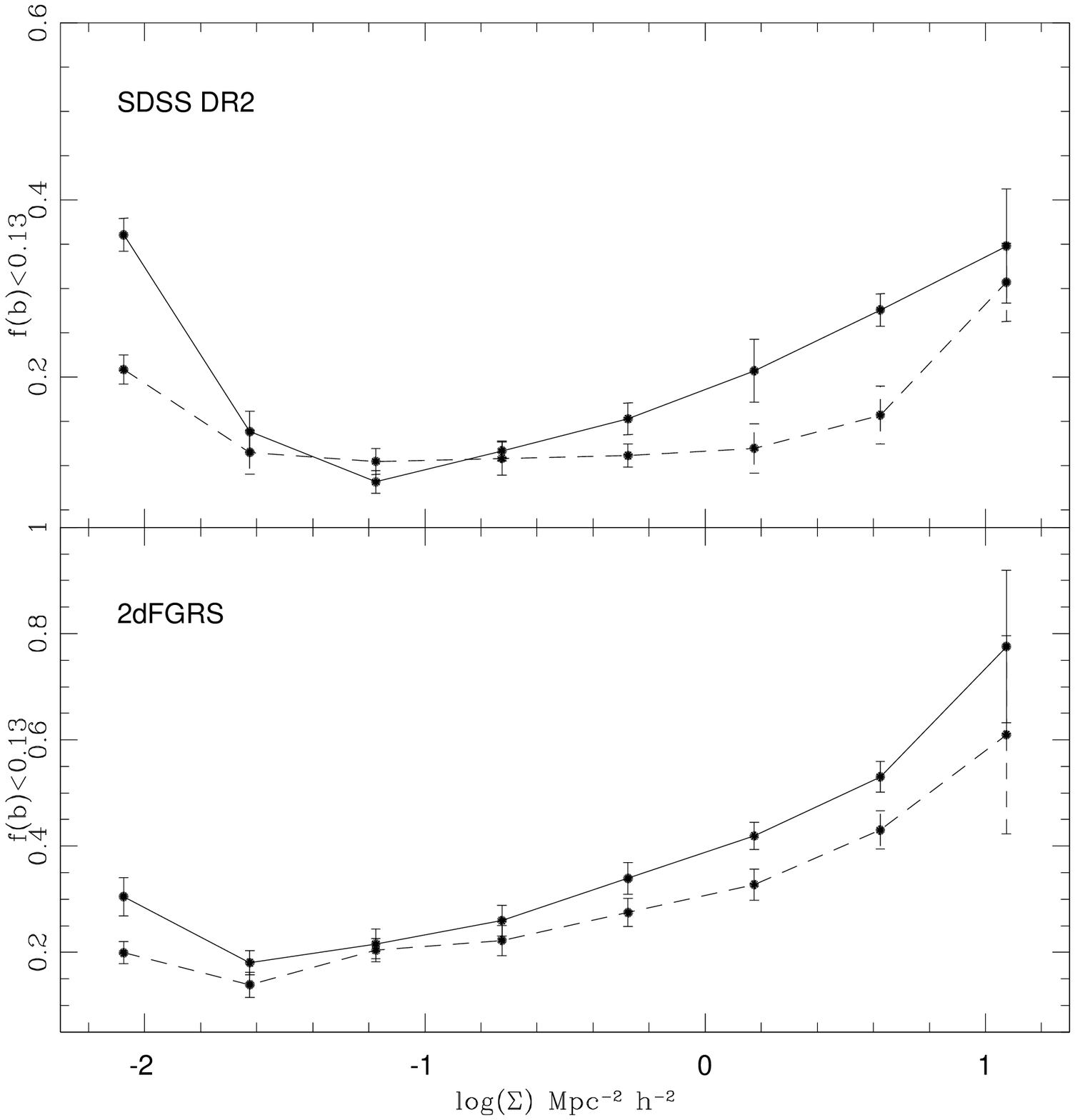,width=9cm,height=9cm}}
\caption{Fraction of low star-forming galaxies (with  $b<0.13$)
as a function of $log$ $ \Sigma$ in close pairs: $r_{\rm p} < 50 \kpc $, 
$\Delta V < 200$ km $s^{-1}$ (solid lines) and in the control samples 
(dashed lines)  in SDSS (upper panel) 
and 2dFGRS (lower panel) surveys. 
}
\label{sfrbsigless}
\end{figure}

In Fig.\ref{sfrbsigless} we show the fraction of galaxies with low star-forming activity
as a function of the projected local density parameter in close pairs and 
in the corresponding control samples. From this figure, it can be appreciated 
the expected dependence on local density for galaxies with and without a close 
companion so that the fraction of low star-forming systems increases with increasing density.
However, we also found an increase of low star-forming galaxies in very low density
regions. For close pairs there is a clear  excess of low star-forming 
galaxies in all environments.

\section {Colours of Galaxies in Pairs}

In this section, we make use of the galaxy colour information in
SDSS-DR2 using Petrosian magnitudes for each object 
to further study galaxies in pairs. 
We calculate the colours with K and extinction corrections following Blanton (2003). 

\begin{figure}
\centerline{\psfig{file=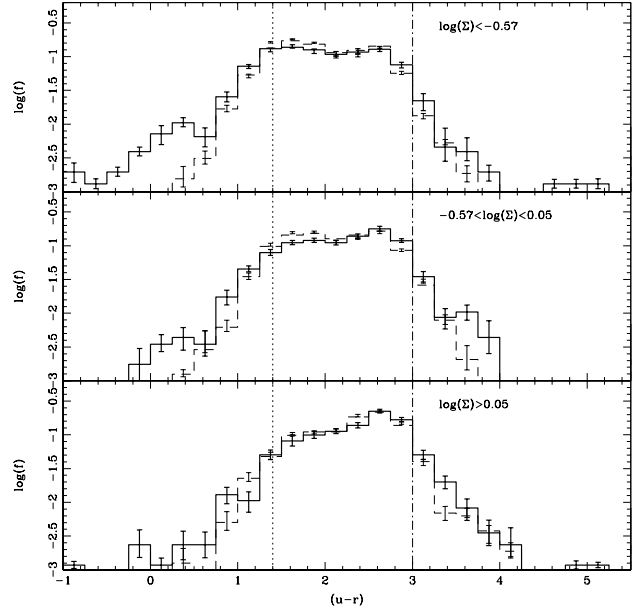,width=9cm,height=9cm}}
\caption{Distribution of $u-r$ colours  for galaxies in close pairs 
($r_p < 50 \kpc $, $\Delta V < 200$ km $s^{-1}$)
(solid line) and in the corresponding control samples (dashed lines),
for the three different environmental classes defined in Section 2.5.
The vertical lines correspond to the adopted blue (dotted) and red (dot-dashed)
limits. Error bars correspond to Poisson standard errors.
}
\label{griz}
\end{figure}

Fig.~\ref{griz} shows the $u-r$ colour distributions for galaxies in close pairs (solid lines) and
in the control sample (dashed lines) for the three environmental classes previously defined.
From this figure it can be appreciated that close pairs have 
distributions with an excess of both red and blue galaxy colours 
with respect to those of the corresponding control sample. This behaviour is more significant in
low density environments. Error bars correspond to Poisson deviations in each colour bin.
A more detailed look yields that, regardless of environment, galaxies in pairs populate more densely 
the colour ranges $u-r < 1.4$ 
(dotted line), corresponding to blue, young stellar populations, and the regions with $u-r > 3$ 
(dot-dashed line) dominated by the more extreme red stellar population, 
than those of the control samples.
A similar analysis for  $g-r$ colours yields a comparable trend with  the blue and the red
extreme defined by  $g-r < 0.4$ and $g-r > 0.95$, respectively.
These values are adopted as colour thresholds to define the extreme red and blue  populations.

\subsection {The extreme blue and red galaxies}

Following the analysis of Section 3,
we also computed the fraction of galaxies with extreme blue and red colour indexes 
as a function of projected separation and relative velocity,
for the same three environmental classes.
We use the colour thresholds determined in the previous section: $u-r < 1.4$ and $g-r < 0.4$ .

\begin{figure}
\centerline{\psfig{file=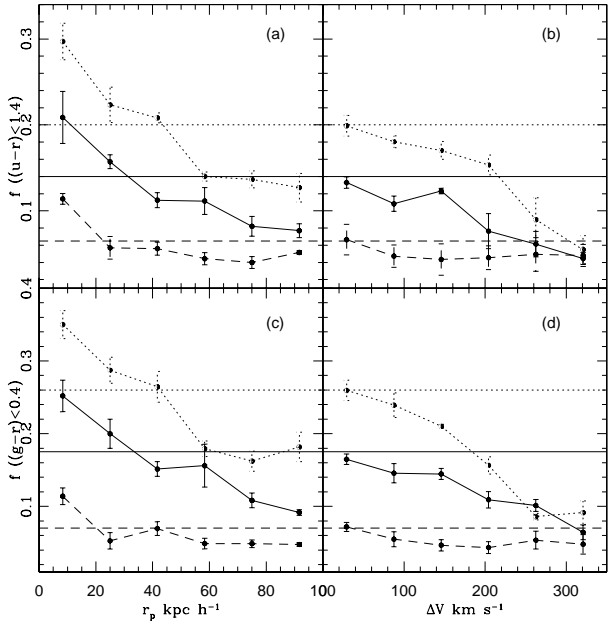,width=9cm,height=9cm}}
\caption{Fraction of galaxies with  $(u-r) < 1.4$ and  $(g-r) < 0.4$
as a function of projected distance, $r_{\rm p}$ (a,c) and relative velocity, 
$\Delta V$ (b,d), for the three different environmental classes defined in Fig.4.
The horizontal lines show to the  fractions of the 
corresponding control samples.
}
\label{colrpV}
\end{figure}

Fig.\ref{colrpV} shows an increase of the fraction of extreme blue galaxies
for small $r_{\rm p} $ and $\Delta V$. 
This behaviour indicates that galaxies in close pairs have an excess of young 
stellar populations with respect to that of the control sample. 
This excess is similar for the three environments (approximately a factor 1.5) although, 
high density regions, systems have to be closer  to reach this excess. The relative separation
thresholds vary from $\approx 50 \kpc $ to $\approx 20 \kpc$ from low to high density regions.
This behaviour is consistent to that found for the star formation activity as it can be 
appreciated from Fig.\ref{SFRrp}.

\begin{figure}
\centerline{\psfig{file=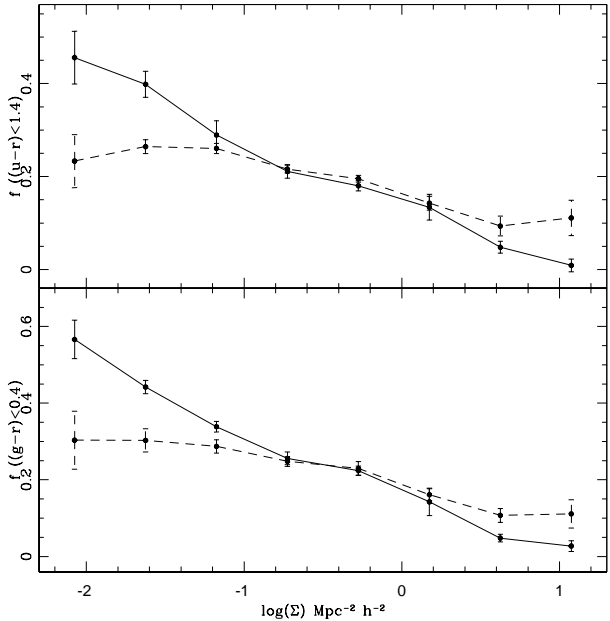,width=9cm,height=9cm}}
\caption{Fraction of galaxies with  $u-r<1.4$ (a) and $g-r<0.4$ (b) 
as a function of projected local density, $\Sigma$, for  close pairs 
(solid lines) and control sample (dashed lines).
}
\label{colblueSig}
\end{figure}

In Fig.\ref{colblueSig} we have plotted the fraction 
of extreme blue galaxies as a function of projected local density 
parameter, $\Sigma$, for systems in both close pairs and control samples. 
As expected, the global trend with $\Sigma$  is complementary to that found 
in Fig.\ref{sfrbsig} so that the fraction of extreme  blue galaxies in  close pairs 
is higher in  low density regions in comparison to 
the corresponding ones of the control sample. This relation inverses for  high densities where 
we find that this fraction  is larger in control sample.
The transition area  ocurs  in a density  range  consistent with that found for 
the fraction of strong star-forming  pairs.

\begin{figure}
\centerline{\psfig{file=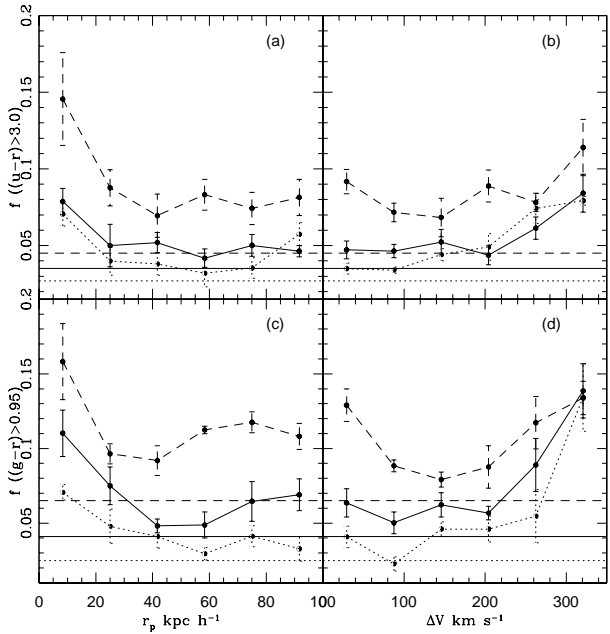,width=9cm,height=9cm}}
\caption{Fraction of galaxies with  $(u-r) > 3.0$ and  $(g-r) > 0.95$
as a function of projected distance $r_{\rm p}$ (a,c) and relative velocity 
$\Delta V$ (b,d) for the three different environmental classes defined in Fig.8.
The horizontal lines show the  fractions  of the 
corresponding control samples. 
}
\label{colredrpV}
\end{figure}

We have also explore the dependence of the fraction of extreme red galaxies on projected 
separation and relative velocity. Fig.\ref{colredrpV} shows an equivalent set 
of plots to that of the extreme blue fraction displayed in Fig.\ref{colrpV},
for the same range of $\Sigma$ values. Note, that the fraction of red galaxies 
in the control samples are almost independent of environment and their boostrap errors
are $\approx 0.005$. 
For galaxies in pairs there is an increase of this fraction
for  small $r_{\rm p}$ and $\Delta V$ values,
 indicating that close pairs have an excess of 
red objects regardless of the environment. The boostrap error bars for the fractions of
the pair and control samples indicate that the signal is statistically meaningful at more than
$3\sigma$-level.

One possible interpretation of this trend is that many  galaxies in pairs have been
very efficient in forming stars at early stages of their evolution so that, currently,
they exhibit red colours. However, from Fig.\ref{SFRrpnonfor} we can appreciate that 
the fractions of low star-forming galaxies decrease for smaller relative separation
for all environments, while the strong star-forming fractions 
increase (Fig.\ref{SFRrp}). The behaviour of colours and star formation
activity  suggests that there is an important fraction of galaxies in very close 
pairs ($r_{\rm p} < 20 \kpc $) which tend to be redder than the rest of the galaxies but
have enhanced star formation activity with respect to the control sample. 
The red colours of these star-forming galaxies in close pairs could be due to obscuration
as the result of dust  stirred up during the encounter which could also hide part of the
star formation activity.

\begin{figure}
\centerline{\psfig{file=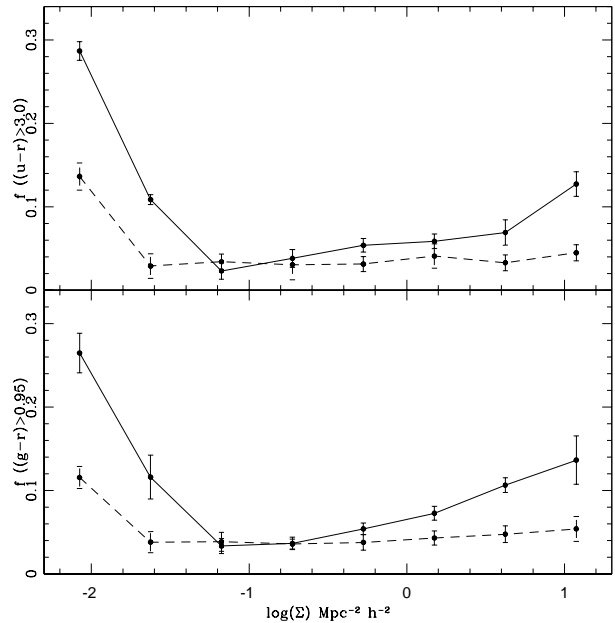,width=9cm,height=9cm}}
\caption{Fraction of galaxies with $u-r>3.0$ (a) and $g-r>0.95$ (b) 
as a function of projected local density, $\Sigma$, in close pairs 
(solid lines) and in the control sample (dashed lines). 
}
\label{colredSig}
\end{figure}

In order to further understand the dependence on environment of the fraction of galaxies with red
colour indexes  in close pairs and in galaxies without a close companion, we have
analysed the fraction of extreme red galaxies in close pairs as a function of the local
density parameter, $\Sigma$. The results are shown in Fig.\ref{colredSig} from where it can be
appreciated  that there is an excess of extreme red galaxies in close  pairs compared to that of 
the control sample in all kind of environment. This trend tends to be stronger in high density 
environments, consistent with the results of low star-forming galaxies discussed in Section 3.

We also notice that at, extremely low density environments, galaxies 
regardless of the presence of a companion, also show a larger fraction of red 
objects in comparison with those of transition density region.
This feature is consistent with the change in the slope of the relation 
between  the fraction of low star-forming galaxies and $\Sigma$
detected for $log$ $\Sigma \le -1.2$ (Fig.\ref{sfrbsigless}).
We argue that these trends can be interpreted as the result of the  growth of small scale 
overdensities in global underdense regions, where the subsequent infall of gas 
and as a consequence, the star formation activity is likely to have been strongly 
reduced at later times. 
Hence generally, galaxies would be less efficiently feeded by gas infall in this regions,
and on top of that, galaxies in pairs would be more efficient at consuming this gas, producing
a larger fraction of red and low star-forming systems.

Previous studies of the colour distribution of galaxies showed an important dependence of the
fraction of red galaxies with environment and luminosity 
(Kauffmann et al. 2004; Balogh et al. 2004; Baldry et al. 2003).
In particular, Balogh et al. (2004)  claimed that the colour distribution is bimodal with only the
relative fraction of galaxies in the red and blue peaks varying with local density and luminosity.
These authors  interpreted their results resorting to   a rapid change of 
galaxy colours due to mergers and interactions. 
Our results support to these findings given the strongest dependence of the star formation activity and
colours of galaxies in pairs on environment with respect to galaxies without a close companion.
 A detail discussion on the bimodal colour distribution of galaxies in pairs will 
be presented in a forthcoming paper.

\section{Conclusions}

We have carried out a detailed analysis of photometric and spectroscopic 
properties of galaxies in pairs at different environments.
The galaxy pair catalogs derived from the 2dFGRS and SDSS-DR2 galaxy surveys available 
comprise 6067 and 6405 pairs, respectively.
In order to characterize environment, we have used a projected galaxy density derived from the 
fifth bright nearest neighbour of each galaxy. We have adopted $M_r<-20.5$ in SDSS and
$b_j < -19.3$ in 2dFGRS so that the derived projected densities of both surveys are nearly
consistent with each other. 
In agreement with previous works our analysis indicates that, globally, 
star formation activity obtained from 2dFGRS and SDSS 
are strong functions of both global environment and orbital parameters.
Our analysis on SDSS pairs confirms previous results found from the 
2dFGRS that the  projected distance $r_{\rm p}<100 \kpc $ and relative
velocity $\Delta V<350$ km $s^{-1}$
are suitable thresholds for the presence of a companion to be correlated 
with effects on the star formation activity of galaxies.

The SDSS galaxy pair catalog has been cleaned of AGNs. 
Conversely, the 2dFGRS may suffer for some contamination by AGNs which can 
affect the relation. This could explain in part some of the small 
differences found in the results from both catalogs. Note, nevertheless, that 
trends are very similar.

Our results found  can be summarized as follows:

\begin{itemize}

\item There is an increase of star formation in pairs for smaller projected separations and
relative velocities in all environments. However, in high density regions galaxies 
have to be closer to statistically show an enhancement respect to galaxies without 
a close companion.

\item The dependence of  the fractions of extremely blue and actively star-forming  galaxies 
in close pairs on local density  show  a different  behaviour respect to
the equivalent fraction of galaxies without a close companion. In low density environment, blue 
and star-forming galaxies tend to be in pairs. In high density regions, galaxies without
a close companion have a higher contribution to the blue and actively star-forming systems. 
The transition local  densities correspond to group environments.

\item  Extremely red and low star-forming galaxies in pairs outnumber those without
a close companion in all environments. 
The fraction of red and low star-forming galaxies without a close companion shows an
increase in the lowest density regions analysed. The corresponding fraction of galaxies in pairs
increases in both the low and the high density ends.
In low density environments,
this effect is more pronounced for pair galaxies owing to  the combined effects of 
the expected lack of gas infall in such regions and the high efficiency of galaxies in pair to 
use the available gas for  new stars at early stages of their evolution.

\item For very close projected distances ($r_{\rm p} < 25 \kpc $ ), there is a statistical 
significant decrease of low star-forming pairs in all environments together with an
important increase of extremely red galaxies in pairs. 
This finding suggests that dust stirred-up during the
encounter may be affecting colours and probably also  obscuring part of the star 
formation activity. 

\end{itemize}

These results show that galaxy-galaxy interactions are an important mechanism for 
triggering star formation regardless of  environment, although at high density systems
have to be closer to react to the presence of a companion. We also found that, 
in low density environments, extremely  blue and star-forming galaxies tend to be 
in pairs, conversely to the situation in high density regions where more violent activity
is found in galaxies without a close companion expect for very close pairs.
This relation has an equivalent behaviour to that of the morphology-density relation
which shows that, in the local universe, actively star-forming systems tend to be located
in low density regions, with a transition area corresponding to group environments.  
Conversely, the red and low star-forming extreme is dominated by galaxies in pairs in 
all environments (except for very close pairs).
 We argue that this trend unveils the ubiquitous effects of interactions
also in previous  stages of evolution which exhausted the gas reservoir in 
close systems. Since these analyses are based on close pairs, contamination by 
interlopers is expected not to be significant.
Dust, however, may be have some effect in very close pairs, although their role is still difficult to 
quantify statistically until an equivalent catalog can be constructed in adequate wavelengths.

\section{Acknowledgments}

This work was partially supported by the
Consejo Nacional de Investigaciones Cient\'{\i}ficas y T\'ecnicas,
Agencia de Promoci\'on de Ciencia y Tecnolog\'{\i}a,  Fundaci\'on Antorchas
and Secretar\'{\i}a de Ciencia y
T\'ecnica de la Universidad Nacional de C\'ordoba.
Funding for the creation and distribution of the SDSS Archive has been provided by the Alfred P. Sloan Foundation, the Participating Institutions, the National Aeronautics and Space Administration, the National Science Foundation, the U.S. Department of Energy, the Japanese Monbukagakusho, and the Max Planck Society. The SDSS Web site is http://www.sdss.org/.

The SDSS is managed by the Astrophysical Research Consortium (ARC) for the Participating Institutions. The Participating Institutions are The University of Chicago, Fermilab, the Institute for Advanced Study, the Japan Participation Group, The Johns Hopkins University, the Korean Scientist Group, Los Alamos National Laboratory, the Max-Planck-Institute for Astronomy (MPIA), the Max-Planck-Institute for Astrophysics (MPA), New Mexico State University, University of Pittsburgh, University of Portsmouth, Princeton University, the United States Naval Observatory, and the University of Washington.

We thank the Referee for a detail revision that
helped to improve this paper.

\label{lastpage}
\end{document}